\documentclass[prl,twocolumn,showpacs,amsmath,amssymb,superscriptaddress]{revtex4-1}

\pdfoutput=1

\newcommand{\zc}[0]{z_\mathrm{c}}

\usepackage{graphicx}% Include figure files
\usepackage{dcolumn}% Align table columns on decimal point
\usepackage{amssymb}
\usepackage{pifont}
\newcommand{\cmark}{\ding{51}}
\newcommand{\xmark}{\ding{55}}
\usepackage{bm}% bold math
\usepackage[usenames,dvipsnames]{color}
\usepackage[colorlinks=true, linkcolor=BrickRed, citecolor=Blue, urlcolor=Blue, filecolor=Blue]{hyperref}
\def\cpc{Comp.\ Phys.\ Comm.}
\def\apj{ApJ}%
\def\apjs{ApJS}%
\def\aap{A\&A}%
\def\jcap{JCAP}%
\def\mnras{MNRAS}%
\def\prd{Phys.\ Rev.\ D}%
\def\prl{Phys.\ Rev.\ Lett.}%
\def\plb{Phys. Lett. B}% 
\def\epjp{Eur.~Phys.~J.~Plus}%
\providecommand{\e}[1]{\ensuremath{\times 10^{#1}}}

\newcommand{\mpl}{M_\mathrm{Pl}}
\newcommand{\impc}{\,\mathrm{Mpc}^{-1}}

\begin{document}

\title{Ultracompact minihalos as probes of inflationary cosmology}

\author{Grigor Aslanyan}
\email{aslanyan@berkeley.edu}
\affiliation{Department of Physics, University of Auckland, Private Bag 92019, Auckland, New Zealand}
\affiliation{Berkeley Center for Cosmological Physics, University of California, Berkeley, CA 94720, USA}

\author{Layne C. Price}
\email{laynep@andrew.cmu.edu}
\affiliation{Department of Physics, University of Auckland, Private Bag 92019, Auckland, New Zealand}
\affiliation{McWilliams Center for Cosmology, Department of Physics, Carnegie Mellon University, Pittsburgh, PA 15213, USA}

\author{Jenni Adams}
\email{jenni.adams@canterbury.ac.nz}
\affiliation{Department of Physics and Astronomy,
University of Canterbury, Christchurch, 8140, New Zealand}

\author{Torsten Bringmann}
\email{torsten.bringmann@fys.uio.no}
\affiliation{Department of Physics, University of Oslo, Box 1048 NO-0316 Oslo, Norway}

\author{Hamish A. Clark}
\email{hamish.clark@sydney.edu.au}
\affiliation{Sydney Institute for Astronomy, School of Physics A28, The University of Sydney, NSW 2006, Australia}

\author{Richard Easther}
\email{r.easther@auckland.ac.nz}
\affiliation{Department of Physics, University of Auckland, Private Bag 92019, Auckland, New Zealand}

\author{Geraint F. Lewis}
\email{geraint.lewis@sydney.edu.au}
\affiliation{Sydney Institute for Astronomy, School of Physics A28, The University of Sydney, NSW 2006, Australia}

\author{Pat Scott}
\email{p.scott@imperial.ac.uk}
\affiliation{Department of Physics, Imperial College London, Blackett Laboratory, Prince Consort Road, London SW7 2AZ, UK}

\date{\today}

\begin{abstract}
\noindent Cosmological inflation generates primordial density perturbations on all scales, including those far too small to 
contribute to the cosmic microwave background. At these scales, isolated ultracompact minihalos of dark 
matter can form well before standard structure formation, if the perturbations have sufficient 
amplitude. Minihalos affect pulsar timing data and are potentially bright sources of gamma rays. 
The resulting constraints significantly extend the observable window of inflation in the presence of cold dark matter,  
coupling two of the key problems in modern cosmology.
\end{abstract}

%\pacs{}

\maketitle

\paragraph{Introduction.---}
Observations of the cosmic microwave background (CMB)
\cite{Hinshaw:2012aka,Ade:2013zuv,Ade:2015xua} provide firm evidence for the
existence of dark matter (DM), as do astrophysical data on galaxy
scales. The same experiments also show that inflation
provides a robust account of the physics of the early
Universe~\cite{Ade:2013rta,Ade:2015lrj}.
However, the microphysical bases of inflation and DM are
unknown and require physics outside the Standard Model.
The leading candidates for DM are weakly-interacting massive particles (WIMPs), which
arise in many well-motivated theories Beyond the Standard Model. Conversely, inflation
typically operates at energies near the scale of grand unified theories~\cite{Liddle:1993ch}.
This \emph{Letter} demonstrates that joint
analyses of the DM and inflationary sectors yield tighter constraints than
those obtained by treating each sector in isolation.

Dark matter and inflation are connected via
primordial density perturbations at small physical scales, which arise
from quantum fluctuations in scalar field(s) during
inflation~\cite{Starobinsky:1980te,*Guth:1980zm,*Linde:1981mu,*Albrecht:1982wi}.
If the amplitude of fluctuations at small scales is significantly larger than at the scales of
the CMB and large scale structure, ultracompact minihalos of DM (UCMHs)  can form shortly after matter-radiation equality~\cite{Berezinsky03,Ricotti09,SS09}.
Recent limits on the UCMH abundance from astrophysical searches for DM annihilation
\cite{SS09,Bringmann11,Lacki10,*JG10}
constrain the power spectrum at scales far smaller than those that contribute to the
CMB.  Limits from pulsar timing~\cite{Clark15a} are projected to lead to similarly strong constraints, and would have the added benefit of not requiring DM to annihilate.
For even larger fluctuation amplitudes,
primordial black hole (PBH) formation is possible~\cite{CarrHawking}, leading to complementary constraints on
inflation~\cite{Peiris:2008be}.

In this \emph{Letter} we provide strong and robust limits on the shape of the
inflationary potential and the primordial power spectrum by combining large-scale
CMB data with small-scale constraints on the number densities of PBHs~\cite{Carr10} and
UCMHs~\cite{Bringmann11,Clark15a,Clark15b}. This method allows one to simultaneously test standard inflation and
the nature of DM,  by cross-correlating the pulsar and
$\gamma$-ray signals. We apply these constraints to a flexible model of inflation, which
can reproduce the results of standard scenarios, \emph{e.g.}\ chaotic~\cite{Linde:1983gd},
hilltop~\cite{Linde:1981mu,Albrecht:1982wi}, and small-field inflation.
Under very conservative assumptions, we find UCMHs provide comparable
constraints on inflation to PBHs, but that they could be even more powerful probes of inflation if we could better understand their formation.

\paragraph{Ultracompact minihalos (UCMHs).---}
A UCMH, as opposed to a regular DM minihalo, collapses before some critical redshift $\zc\gtrsim\mathcal{O}(100)$. These halos form in isolation,
with extremely small velocity dispersions, via almost pure radial infall. This produces a steep density profile $\rho_{DM}\varpropto r^{-9/4}$~\cite{Ricotti09,Fillmore84,*Bertschinger85,*Vogelsberger09} with an inner plateau due to 
finite DM angular momentum~\cite{SS09,Bringmann11} and possible DM
self-annihilation.  This compact core makes UCMHs insensitive
to tidal disruption~\cite{Berezinsky2006,*Berezinsky2008,*Berezinsky12,*Berezinsky13}. 
Because annihilation scales with $\rho_{DM}^2$, they 
are excellent indirect DM search targets~\cite{SS09,Bringmann11,Lacki10,*JG10}.  
%Microlensing searches appear promising~\cite{Ricotti09,Li12}, but are currently not sensitive enough to provide competitive limits; the same is true for pico- and femtolensing.
%Time-delay lensing offers better prospects,
Time-delay lensing can constrain the UCMH number density,
as a UCMH that passes near the line of sight between Earth and a distant pulsar 
would cause a change in its observed pulsation rate~\cite{Clark15a}.  

Assuming that UCMHs track the bulk DM density, both on cosmological and Galactic scales, limits on their cosmological abundance can be inferred from local limits on the UCMH number density.
If DM annihilates, $\gamma$-ray limits from \textit{Fermi}-LAT
provide the strongest bounds~\cite{Bringmann11}.
The impacts of WIMP annihilation in UCMHs on reionisation may also be apparent in the
CMB~\cite{Zhang11,*Yang11a,Yang11b}.
Constraints from pulsars~\cite{Clark15a}, based on gravitational effects only, would be entirely model-independent~\footnote{That is, except for 
DM models with an intrinsic cutoff in the power spectrum at scales larger
than probed by pulsar timing data.}; while extending over a smaller range of 
scales, projected limits are at least as constraining as gamma-ray constraints.
Complementary but weaker constraints can also be obtained from CMB spectral 
distortions~\cite{Chluba:2012we,Chluba:2013pya}.
These limits constrain the processes that could have 
formed UCMH-seeding overdensities in the early Universe~\cite{Bringmann11,Li12,Shandera12,Lacki10,*JG10,Yang12,*Berezinsky2011,Anthonisen,Yang11b,Clark15b}.

The fraction of DM in UCMHs with present-day mass $M_0$ is $f = \Omega_{\rm
UCMH}/\Omega_\chi = (M_0/M_i)\beta(R)$ \cite{Bringmann11}, where ${M_i}$ is the initial
mass contained in an overdense region of comoving size $R$. For 
a Gaussian distribution, the fraction of 
perturbations that collapse to form UCMHs is 
\begin{align}
 \beta(R)=\frac{1}{\sqrt{2\pi}\sigma_{\chi,\mathrm{H}}(R)}\int_{\delta_\chi^\mathrm{min}}^{\delta_\chi^\mathrm{max}}\!\!\exp\!\left[-\frac{\delta_\chi^2}{2\sigma_{\chi,\mathrm{H}}^2(R)}\right]\,{\rm d}\delta_\chi\,.
 \label{betafull}
\end{align}
Here, the minimum density contrast $\delta_{\rm min}$ required for UCMH formation is the 
minimum amplitude at horizon entry that a perturbation must possess for it to have sufficient 
time to begin nonlinear collapse before $\zc$.  
Typically $\delta_{\rm min}\sim10^{-3}$  \cite{Ricotti09,Bringmann11}.
%the exact value is roughly proportional to $\zc$, and depends logarithmically on the perturbation scale $k$ \cite{Bringmann11}.
If the initial overdensity is too large, $\delta_\chi\geq \delta_\chi^\mathrm{max}\sim\mathcal{O}(1)$, 
a PBH rather than a UCMH would form.  However,
since $\delta_\chi^\mathrm{min}\ll  \delta_\chi^\mathrm{max}$,
$\beta(R)$ is independent of $ \delta_\chi^\mathrm{max}$ to a very good approximation.
The quantity $\sigma_{\chi,\mathrm{H}}(R)$ is the mass variance of perturbations at the
time $t_{k_R}$  of horizon-entry of the scale ${k_R}\sim 1/R$. It is roughly proportional to
the total size of perturbations at  $t_{k_R}$,
 $\sigma^2_\mathrm{H}(R)=A_\chi^2(k_R)\,\delta_\mathrm{H}^2(t_{k_R})$,
where the factor $A_\chi$ depends on the initial spectrum of
perturbations produced during inflation and the expansion history 
since~\cite{Bringmann02,*Blais03}.
In the special case of an almost scale-free spectrum with a spectral index $n_s(k)$ that runs only at
first order~\cite{Bringmann11},
\begin{align}
 \label{alpha}
 A_\chi^2(k) &=& \frac{9}{16}\int_0^\infty \!\!dx\,x^{n_s(k_*)+2+\alpha_s \ln\left(\frac{xk}{k_*}\right)}\left(\frac{k}{k_*}\right)^{\alpha_s \ln x} \nonumber\\
       &&\times\ W^2_{\rm TH}(x) \frac{T_\chi^2\left(x/\sqrt{3}\right)} {T_\mathrm{r}^2\left(1/\sqrt{3}\right)}\,,
\end{align}
where $W_{\rm TH}$ is the Fourier transform of a spherical top-hat window function, $T_\mathrm{r}$ ($T_\chi$) is the radiation (DM) transfer function,
and $\alpha_s \equiv \mathrm{d}n_\mathrm{s}/\mathrm{d}\ln k$ is the running of the spectral index $n_s$.
However, inflationary models generally have a scale dependence beyond $\alpha_s$ and we 
therefore apply UCMH constraints using the \textit{local} slope of the power spectrum instead, 
i.e.~we set $\alpha_s=0$ and replace $n_s(k_*)\to n_s(k)$.

The most crucial non-primordial parameter for the
UCMH abundance is $\zc$, the lowest redshift at which collapse happens
radially and in full isolation.
Smaller $\zc$ allows smaller-amplitude perturbations to form
UCMHs, as perturbations have longer to collapse.
This parameter is poorly
constrained, as it represents the redshift at which the approximations of spherical collapse
and secondary infall
break down~\cite{Fillmore84,*Bertschinger85,*Vogelsberger09}. These are excellent
approximations at $z\gtrsim1000$,
but
when nonlinear structure formation begins at $z\lesssim30$, these conditions certainly do not hold.
%and spherical collapse is not valid.
In this \emph{Letter} we use $z_c=1000$ as an extremely conservative
choice, but show how limits improve with $\zc=500$ and $\zc=200$, which are both
 realistic possibilities.

\paragraph{Limits on the UCMH abundance.---}
Gamma-ray fluxes depend on $\rho_{DM}$, the
DM mass $m_\chi$, annihilation cross-section $\langle\sigma v \rangle$, and annihilation
branching fractions into different final states. Lighter WIMPs produce larger fluxes;
we make the conservative choice $m_\chi=1$\,TeV. We assume an NFW profile for the Milky Way,
the canonical `thermal value' for the annihilation cross-section
$\langle\sigma v \rangle=3\times10^{-26}$\,cm$^3$\,s$^{-1}$, and 100\%
annihilation into $b\bar b$ pairs (which produce $\gamma$-rays mostly by neutral pion
decay). The limits are not especially sensitive to these
assumptions~\cite{SS09,Bringmann11}. We adopt the likelihood function of
Refs.~\cite{Bringmann11, Shandera12} for the abundance of UCMHs indicated
by \textit{Fermi}-LAT $\gamma$-ray observations~\cite{2FGL}, based on the diffuse flux
from the Galactic poles, and the non-observation of DM minihalo sources in the first
year of all-sky survey data.

If DM does not annihilate, pulsars provide the only realistic means of detecting low-mass UCMHs. Here we apply the projected constraints from the individual-halo Shapiro delay detection method of Ref.~\cite{Clark15a}, assuming a transit detection threshold of 20 ns. Assuming non-detection of UCMH transits within 30-year pulsar timing data provides the strongest projected gravitational bound on UCMHs with masses $\sim 10^{-3} M_\odot$. While the assumed detection threshold provides relatively weak limits on the fraction of DM contained within UCMHs compared to those from gamma-ray searches, it may soon be improved with the development of high-sensitivity pulsar timing arrays, improved understanding of the nature of pulsar timing noise, and increased observation time in existing millisecond pulsar surveys.
The corresponding limits on the power spectrum only apply in the local vicinity of the
scale $k_R$, \emph{i.e.}, where the predicted power spectrum is
approximately locally power law.
Although pulsar limits are weaker than $\gamma$-ray ones, they are
purely gravitational, and would apply regardless
of the precise particle properties of DM.

\paragraph{The observable window of inflation.---}
We use a phenomenological inflation
model that can mimic many plausible scenarios, including large-field and
small-field inflation, which have large and small values of the tensor-to-scalar ratio $r$,
respectively. We parametrize the inflationary potential as
\begin{align}
 \label{eqn:V}
 V(\phi) = \sum_{n=0}^4 \frac{V_n }{n!} \, \left(\phi - \phi_*\right)^n,
\end{align}
where $\phi_*$ is the inflaton field value when the pivot
scale $k_* = 0.05\impc$ leaves the horizon, which is fixed to $\phi_* = 0$ without loss of
generality. The real constants $V_n$ are related to the
slow-roll parameters $\{\epsilon_*, \eta_*, \xi_*, \omega_* \}$ evaluated at $\phi=\phi_*$ by
\begin{align}
 V_1 &= \frac{V_0 \sqrt{2 \epsilon_*}}{\mpl}      & & \text{ with } &  \epsilon_* &= \frac{\mpl^2}{2} \left( \frac{ V'}{V} \right)^2 , \\
 V_2 &= \frac{V_0 \, \eta_*}{\mpl^2}           & & \text{ with } & \eta_*   &= \mpl^2 \frac{V''}{V}, \\
 V_3 &= \frac{\xi_*^2V_0^2}{\mpl^4V_1}    & & \text{ with } & \xi_*^2  &= \mpl^4 \frac{V' V'''}{V^2}, \\
 V_4 &= \frac{\omega_*^3V_0^3}{\mpl^6V_1^2}  & & \text{ with } & \omega_*^3 &= \mpl^6 \frac{V'^2 V''''}{V^3},
 \label{eqn:SR_params}
\end{align}
and $V_0 =V(\phi_*)$, where $\mpl^2 = 1/8 \pi$.

Expanding $V$ to fourth-order in $\phi$ allows the primordial spectrum $\mathcal P_\zeta(k)$ to have a running spectral index
$\alpha_s $ and a higher order running-of-the-running
$\alpha_s ' \equiv d \alpha_s/d\ln k$, giving significant freedom in the
shape of $\mathcal P_\zeta(k)$, although this cannot easily replicate $V(\phi)$ with a step or sinusoidal oscillations.
The potential~\eqref{eqn:V}
was used in Refs.~\cite{Lesgourgues:2007gp,Planck:2013jfk,*Ade:2015lrj} as an
empirical description of the primordial epoch, constrainable in a CMB
``observable window'' of scales $10^{-6}  \lesssim k / \impc \lesssim 10^{-1}$.   Measurements of the power spectrum put tight limits on the slow-roll parameters, ensuring the plausible domain of validity of \eqref{eqn:V} is larger than $\mpl$, and therefore describes the potential through $\mathcal O(10-100)$ $e$-folds of inflation. Furthermore, \eqref{eqn:V} is the minimal polynomial potential for which $n_s$, $\alpha_s$, and $\alpha_s'$ are independent and potentially non-trivial.

Using the \textsf{ModeCode} inflation package~\cite{Mortonson:2010er,*Easther:2011yq,*Norena:2012rs,*Price:2014xpa},
we solve the equations of motion for $\phi(t)$ and the perturbations
$\delta \phi(t,k)$ numerically, assuming the Bunch-Davies initial condition
on sub-horizon scales~\cite{Bunch:1978yq}.
We do not require slow-roll to hold
during inflation or
 $V>0$ except at $V_0$, since inflation must end before $V<0$.
We also include results using the inflation module from
\textsf{Class}~\cite{Lesgourgues:2011re,*Blas:2011rf}, which replicates previous techniques~\cite{Lesgourgues:2007gp,Planck:2013jfk,*Ade:2015lrj}. We find no difference between the two implementations where they overlap.

For fixed $V_n$ the number of $e$-folds $N_*$ between horizon exit for
the pivot scale $k_*$ and the end of inflation, as well as
the primordial power spectrum parameters $A_s$, $n_s$, $\alpha_s$, and $\alpha_s'$ and the tensor-to-scalar ratio $r_{0.002}$ at the alternate scale of $k=0.002\impc$, are derived parameters.

\paragraph{UCMH constraints on inflation.---}
Including UCMHs and PBHs increases the highest constrainable wavevectors in
$\mathcal P_\zeta (k)$ to $k \sim 10^{18} \impc $, significantly extending the
range $\Delta \phi$ over which $V(\phi)$ can be
reconstructed.  While the UCMH limits on $\mathcal P_\zeta(k)$ at
these small scales are orders of magnitude less severe than in the CMB range,
including them has a strong effect on the higher order runnings in the spectrum.
For identifying successful inflationary solutions, we require that all modes $k \le 10^{18} \impc$ leave the horizon during inflation, corresponding
to $N_* \gtrsim 45$. We assume inflation can end by a hybrid
transition or some other mechanism not necessarily captured in Eq.~\eqref{eqn:V}.

\begin{table}
\renewcommand{\arraystretch}{1.5}
\setlength{\tabcolsep}{6pt}
\begin{tabular}{c | c c|c c|c c |c  c}
Scan \# & 0    & 1    & 2    & 3    & 4    & 5    & 6    & 7     \\
  \hline
CMB    & \cmark & \cmark & \cmark & \cmark & \cmark & \cmark & \cmark & \cmark  \\
$\gamma$-ray & \xmark       & \xmark       & \cmark       & \xmark       & \cmark       & \xmark       & \cmark       & \xmark        \\
Pulsar & \xmark & \xmark & \xmark & \cmark & \xmark & \cmark & \xmark & \cmark  \\
 PBH    & \xmark & \cmark & \xmark & \xmark & \xmark & \xmark & \xmark & \xmark  \\
  $z_c$& \multicolumn{2}{c|} {---}  & \multicolumn{2}{c|} {1000} & \multicolumn{2}{c|} {500}  & \multicolumn{2}{c} {200}
\end{tabular}
\caption{Scan specifications. The rows show when we use CMB, $\gamma$-ray UCMH, (projected) pulsar UCMH, and PBH data, and the redshift $z_c$ for UCMH formation.}
\label{scan_table}
\end{table}

%\begin{table}
%\renewcommand{\arraystretch}{1.5}
%\begin{tabular}{c | c|c|c|c|c}
%Scan & CMB & $\gamma$-ray & Pulsar & PBH & $z_c$\\
%\hline
%0 & \cmark & \xmark & \xmark & \xmark & --- \\
%1 & \cmark & \xmark & \xmark & \cmark & --- \\
%\hline
%2 & \cmark & \xmark & \cmark & \xmark & 1000 \\
%3 & \cmark & \cmark & \cmark & \xmark & 1000 \\
%\hline
%4 & \cmark & \xmark & \cmark & \xmark & 500 \\
%5 & \cmark & \cmark & \cmark & \xmark & 500 \\
%6 & \cmark & \cmark & \cmark & \cmark & 500 \\
%\hline
%7 & \cmark & \xmark & \cmark & \xmark & 200 \\
%8 & \cmark & \cmark & \cmark & \xmark & 200 \\
%9 & \cmark & \cmark & \cmark & \cmark & 200 \\
%\end{tabular}
%\caption{Scan specifications. The columns show when we use CMB, $\gamma$-ray UCMH, pulsar UCMH, and PBH data, and the redshift $z_c$ for UCMH formation.}
%\label{scan_table}
%\end{table}

\begin{figure*}
\includegraphics[height=0.89\columnwidth]{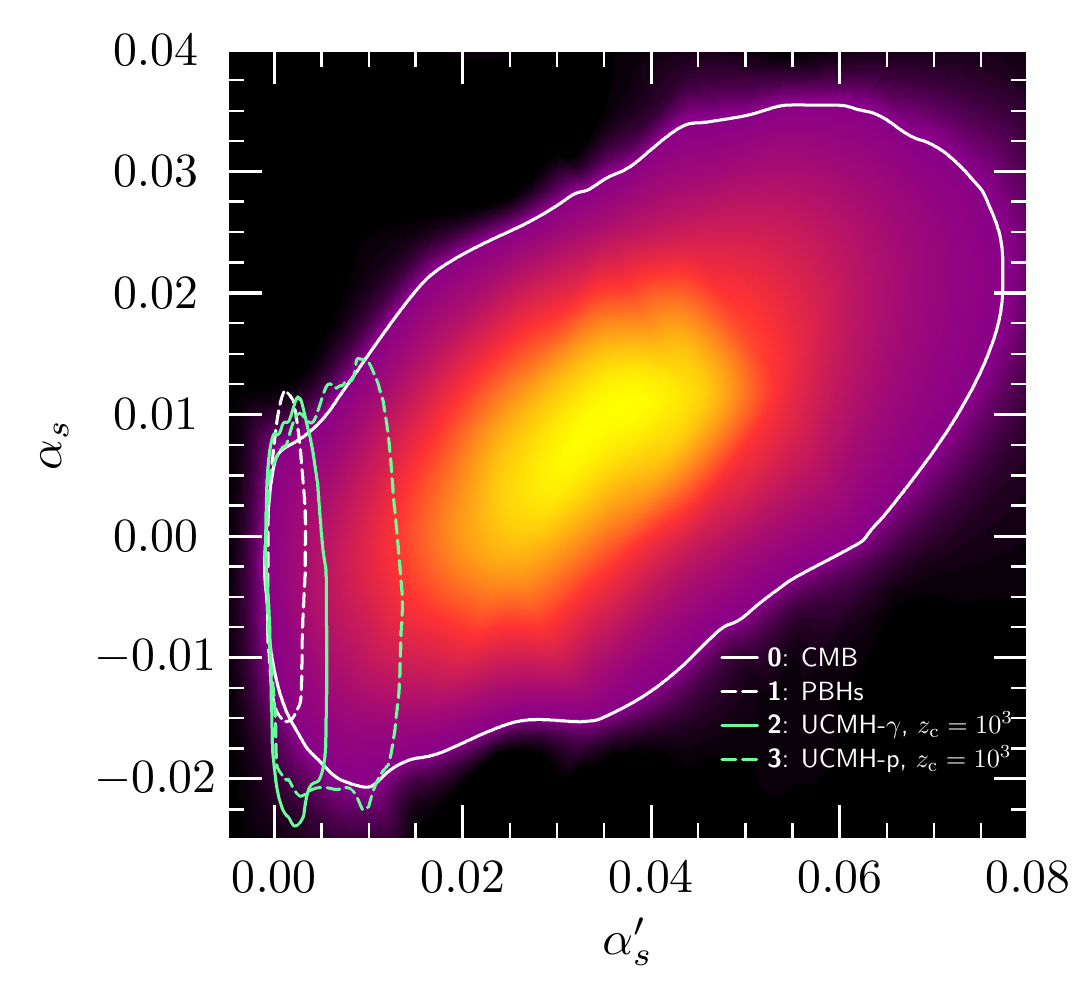}
\includegraphics[height=0.89\columnwidth]{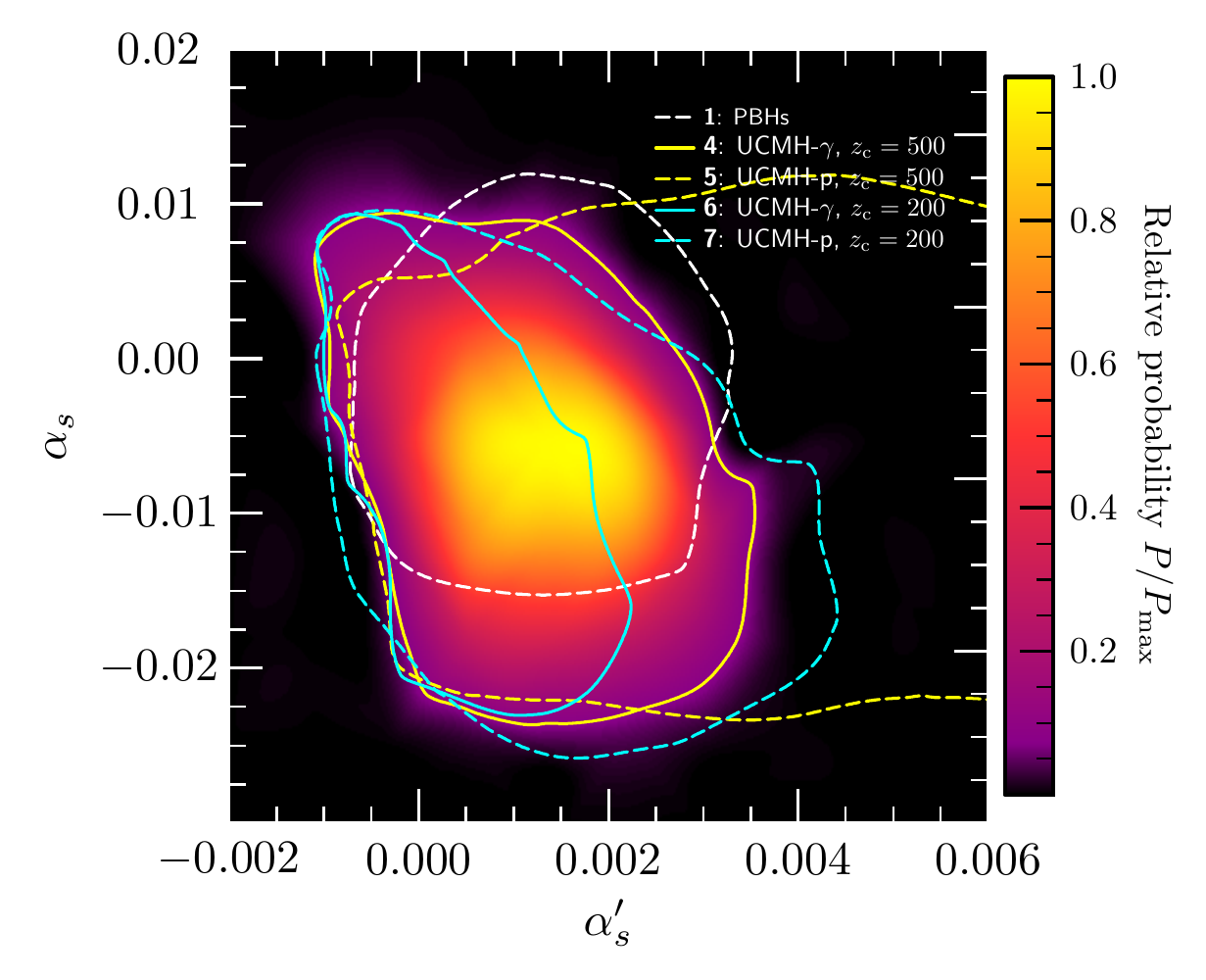}
\includegraphics[height=0.89\columnwidth]{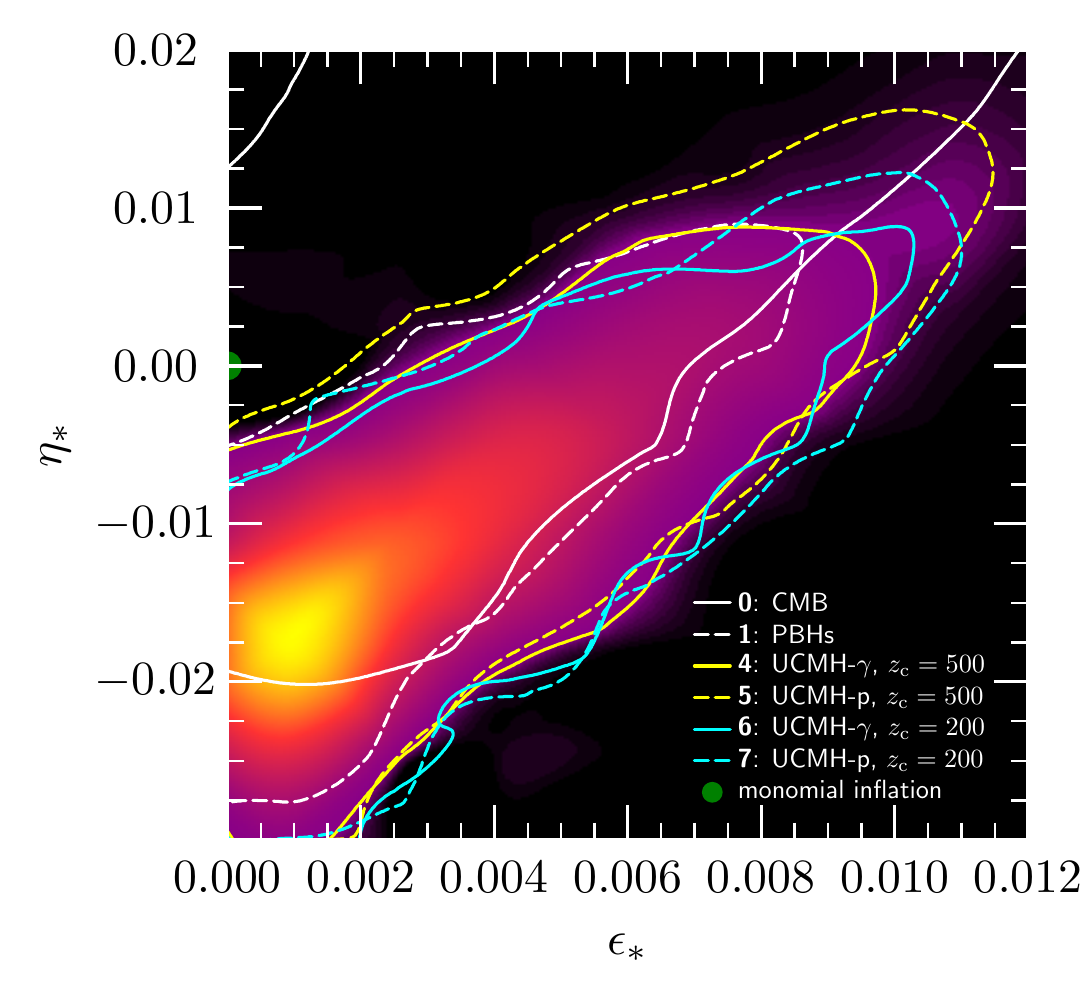}
\includegraphics[height=0.89\columnwidth]{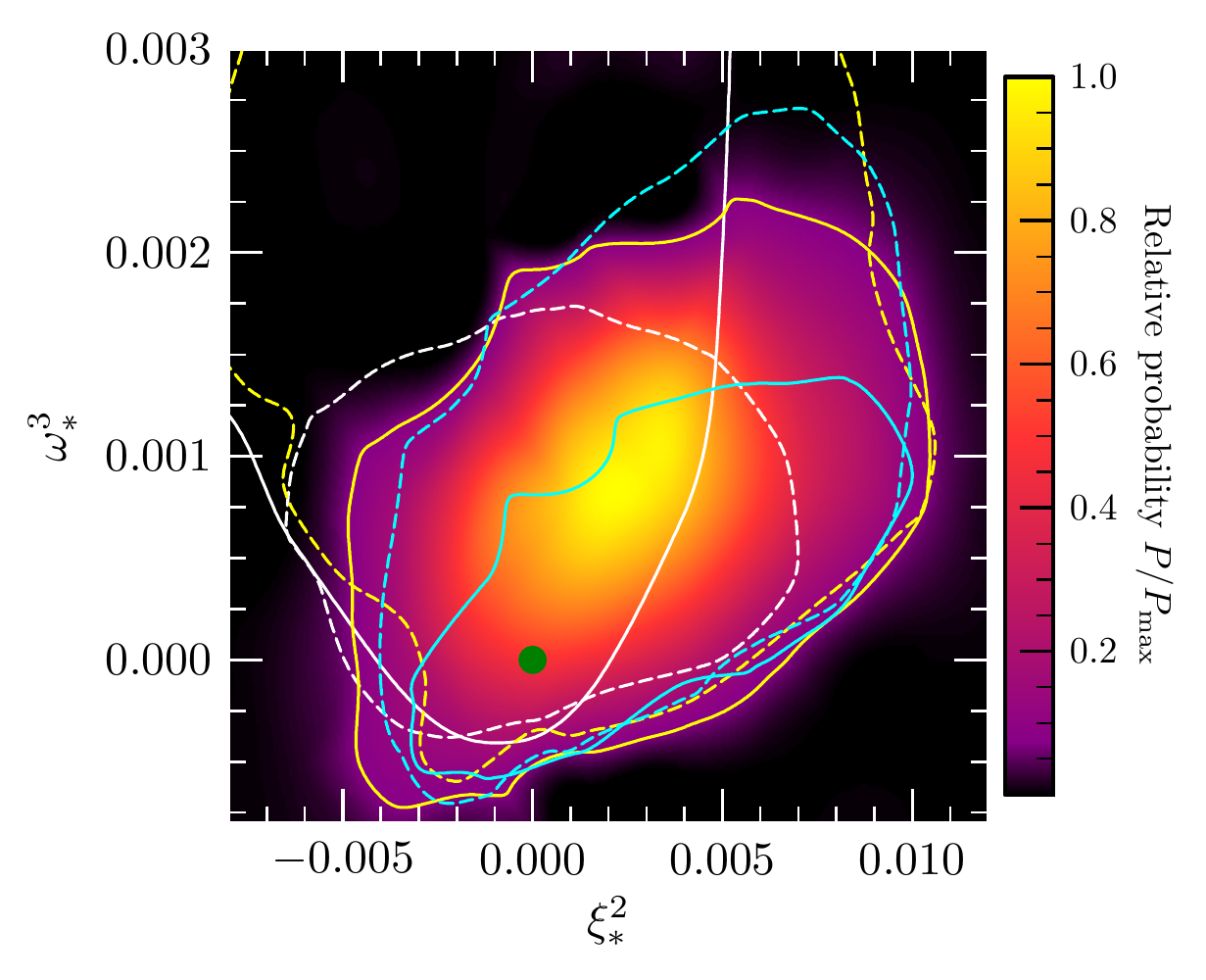}
\caption{(\emph{Top row}) $95\%$ credible regions (CRs) for the running $\alpha_s\equiv d\,n_s/d\,\ln k$ and the running-of-the-running $\alpha_s^\prime\equiv d^2\,n_s/d\,\ln k^2$ of the
primordial power spectrum at the pivot scale $k_*=0.05\impc$. Curves correspond to
different combinations of data. `UCMH-p' and `UCMH-$\gamma$' refer to pulsar and $\gamma$-ray constraints on UCMHs, respectively.
Numbers in legends refer to scans
in Table~\ref{scan_table}.  The left and right panels are shaded by the posterior pdfs of
Scans 0 and 5, respectively.
(\emph{Bottom row})  $95\%$ CRs for the inflationary slow-roll parameters, shaded by the
posterior pdf of Scan 5. The green dot shows predictions of monomial models.}
\label{fig2}
\end{figure*}

\begin{table}
\renewcommand{\arraystretch}{1.5}
\setlength{\tabcolsep}{6pt}
\begin{tabular}{c|c|c|c}
  & \text{Scan }0 & \text{Scan }2 & \text{Scan }6 \\
\hline
$n_s$ & $0.960^{+0.011}_{-0.011}$ & $0.9650^{+0.0104}_{-0.0094}$ & $0.9650^{+0.0101}_{-0.0097}$ \\
$\alpha_s$ & $0.008^{+0.020}_{-0.020}$ & $-0.006^{+0.014}_{-0.014}$ & $-0.008^{+0.014}_{-0.012}$ \\
$\alpha_s^\prime$ & $0.035^{+0.037}_{-0.029}$ & $0.0025^{+0.0024}_{-0.0027}$ & $0.0005^{+0.0013}_{-0.0012}$ \\
$r_{0.002}$ & $<0.28$ & $<0.14$ & $< 0.12$
\end{tabular}
\caption{$95.5\%$ CIs for the primordial parameters from the CMB-only (Scan 0), compared
  to conservative UCMH likelihoods (Scan 2) and the tighter constraints
  from UCMH with smaller $z_c$ (Scan 6).
}
\label{limits_table}
\end{table}

We obtain posterior probabilities for the primordial spectra and inflationary parameters
using the \textsf{Cosmo++} package~\cite{Aslanyan:2013opa} and the nested
sampling code \textsf{MultiNest}~\cite{Feroz:2007kg,*Feroz:2008xx,*Feroz:2013hea} (plotted with \textsf{pippi}~\cite{pippi}). We use the \emph{Planck} 2015
TT,TE,EE+lowP likelihood code~\cite{Aghanim:2015xee} and the \emph{Fermi}-LAT
and pulsar UCMH likelihoods described above.
We compute $\gamma$-ray and pulsar likelihoods for
$10^{-6} < k/\impc < 10^{18}$, applying at each $k$ the correction for the local
slope of the power spectrum described in Appendix B3 of
Ref.~\cite{Bringmann11};
finally selecting the $k$ that produces the strongest constraint.

We use uniform priors for the cosmological parameters $\Omega_\mathrm{b}h^2$,
$\Omega_\mathrm{c}h^2$, $h$, and $\tau$, and for the slow-roll parameters
$\epsilon_*$, $\eta_*$, $\xi_*^2$, and $\omega_*^3$, with a log prior for the ratio
$V_0/\epsilon_* \varpropto A_s$, matching previous analyses~\cite{Lesgourgues:2007gp,Ade:2015lrj}.

We perform several scans with different assumptions (Table~\ref{scan_table}).
The fiducial Scan 0 uses only CMB data and
agrees well with the \emph{Planck} analysis~\cite{Ade:2015lrj}.
Scan 1 adds PBH constraints, employing a step-function likelihood from the
implementation of the limits of Ref.~\cite{Carr10} in \textsf{DarkSUSY}~\cite{Anthonisen}, following Ref.~\cite{Bringmann11}. Different
scans use different UCMH parameters: $z_c=1000$ (Scans 2--3), $z_c=500$
(Scans 4--5) or $z_c=200$ (Scans 6--7). Scans 2, 4, and 6 add only UCMH
constraints from $\gamma$-rays, while   Scans 3, 5, and 7 use projected pulsar limits instead.
Table~\ref{limits_table} shows the 95\% credible intervals (CIs) for the primordial parameters for three scans.

Fig.~\ref{fig2}
shows the $95\%$ CIs for $\alpha_s$ and $\alpha_s'$.
Compared to the CMB alone, using small-scale data (Scans 1--7) significantly
tightens the credible regions on all the primordial parameters, severely limiting
the shape of the inflationary potential.
The 95\% CI for the running-of-the-running is 
$0 \lesssim \alpha_s' \lesssim 0.05$ (Scan 2) or
$-1\e{-3}\lesssim\alpha_s'\lesssim2\e{-3}$ (Scan 6), implying the non-observation
of DM structures can robustly constrain the highest-order derivatives
of $\mathcal P_\zeta(k)$.  The posteriors depend strongly on
$\zc$, with much tighter constraints for $z_c=200$ than for $z_c=1000$. The
UCMH likelihoods alone produce similar results to PBHs,
but only become truly competitive with PBHs for $z_c\lesssim500$, while
the combination of PBHs and UCMHs with $z_c=200$ can constrain cosmological
parameters much more tightly than either UCMHs or PBHs alone. More detailed
knowledge of $z_c$ will thus be instrumental in drawing tight constraints on primordial
parameters from UCMHs.

Fig.~\ref{fig2} also shows $95\%$ CIs for the inflationary parameters.  Comparing to the
\emph{Planck} results,
the first two slow-roll parameters have a much narrower range, 
$\epsilon_* \lesssim 0.009$ and $-0.025 \lesssim \eta_* \lesssim 0.01$.  Scans 1--7
prefer inflation with a lower value of the tensor-to-scalar ratio,
$r \lesssim 0.13$, compared to $r \lesssim 0.28$ at 95\% CI for Scan 0,
even though small-field inflation ($\epsilon_* < \eta_*$) is not given equivalent weight to large-field inflation due to uniform priors on $\epsilon_*$ and $\eta_*$~\cite{Adshead:2008vn}.
Scan 6
($z_c=200$) has the tightest contours for the inflationary parameters,
with $r \lesssim 0.12$.
Including BICEP2/\emph{Keck Array} CMB polarization
data~\cite{Ade:2015tva} might further reduce $r$.
The higher-order slow-roll parameters $\xi_*^2$ and $\omega_*^3$ are pushed
significantly toward zero by the DM constraints, mirroring the reduced
range of $\alpha_s'$ in Fig.~\ref{fig2}.  
For comparison with some concrete models, we also show the predictions of a simple potential $V=\lambda\phi^n$.

We have also artificially weakened the limit on $\mathcal P_\zeta$ from UCMH constraints (not plotted) by a factor of $\sim10$, finding little change in the results, as most scenarios predict $\alpha_s^\prime > 0$ and are ruled out by even weak limits on smaller scales.

\paragraph{Conclusion.---}
Searches for UCMHs are sensitive to a wide range of amplitudes and slopes in the
primordial power spectrum. UCMHs can thus be used to directly
probe the preferred parameter region in inflationary models, in a way
complementary to the CMB.
Under conservative assumptions about the particle nature of dark matter, pulsar timing observations alone will be
able to exclude a large portion of the otherwise-viable
region of inflationary parameter space. If DM annihilates, non-observation of
$\gamma$-rays from DM point sources by \textit{Fermi} already imposes 
tight constraints.

We have demonstrated for the first time that even a conservative application of the current
understanding of the formation and evolution of UCMHs leads to significant limits on
inflation. Future analyses would benefit from improved understanding of UCMH formation,
particularly the minimum collapse redshift $z_c$ at which a halo can be considered a UCMH
that is not significantly affected during the epoch of non-linear structure
formation. Given the strength of the limits
when we assume
$z_c\lesssim500$, urgent investigation is needed into the formation and gravitational history of the
earliest bound objects in the Universe.

\begin{acknowledgments}
\paragraph{Acknowledgments.---}
LCP is supported in part by the Department of
Energy under grant DESC0011114.  PS is supported by STFC (ST/K00414X/1 and ST/N000838/1). HAC is supported by the University of Sydney through an Australian Postgraduate Award (APA).  We acknowledge the use of the New Zealand
eScience Infrastructure (NeSI) high-performance computing facilities, which are funded
jointly by NeSI's collaborator institutions and through the Ministry of Business, Innovation
\& Employment's Research Infrastructure programme [{\url{http://www.nesi.org.nz}}].

\end{acknowledgments}

\bibliography{refs}

\end{document}